\documentclass[10pt,a4paper,oneside,openright]{article}
\usepackage[reqno,sumlimits]{amsmath}
\usepackage{latexsym}
\usepackage{amsmath}
\usepackage{graphics}
\usepackage{amsfonts}
\usepackage{amssymb}
\usepackage{theorem}
\usepackage{amsbsy}
\usepackage{color}
\usepackage{epsfig}
\usepackage{lscape}
\usepackage{graphicx}
\usepackage{wrapfig}
\usepackage[footnotesize,bf]{caption}
\usepackage{verbatim}
\usepackage{subfig}
\usepackage[absolute]{textpos}

\title{\bf{The resistance of randomly grown trees}}
\author{E.R. Colman and G.J. Rodgers \\ \emph{Department of Mathematical Sciences,} \\ \emph{Brunel University, Uxbridge, Middlesex UB8 3PH, U.K.}}

\begin{document}
\nocite{*}
\begin{textblock}{14.5}(0.5,0.2)
\noindent \footnotesize This is an author-created, un-copyedited version of an article accepted for publication in Journal of Physics A. IOP Publishing Ltd is not responsible for any errors or omissions in this version of the manuscript or any version derived from it. The Version of Record is available online at doi:10.1088/1751-8113/44/50/505001.
\end{textblock}

\maketitle
\begin{abstract}
An electrical network with the structure of a random tree is considered: starting from a root vertex, in one iteration each leaf (a vertex with zero or one adjacent edges) of the tree is extended by either a single edge with probability $p$ or two edges with probability $1-p$. With each edge having a resistance equal to $1$, the total resistance $R_{n}$ between the root vertex and a busbar connecting all the vertices at the $n^{th}$ level is considered. Representing $R_{n}$ as a dynamical system it is shown that $\langle R_{n} \rangle$ approaches $(1+p)/(1-p)$ as $n\rightarrow\infty$, the distribution of $R_{n}$ at large $n$ is also examined. Additionally, expressing $R_{n}$ as a random sequence, its mean is shown to be related to the Legendre polynomials and that it converges to the mean with $|\langle R_{n}\rangle-(1+p)/(1-p)|\sim n^{-1/2}$.
\end{abstract}

\section{Introduction}
\begin{figure}[h]
  \centering
  \subfloat[]{\label{fig:gull}\includegraphics[width=0.18\textwidth]{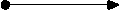}}\hspace{1cm}
  \subfloat[]{\label{fig:tiger}\includegraphics[width=0.18\textwidth]{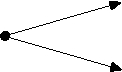}}\hspace{1cm}
  \subfloat[]{\label{fig:mouse}\includegraphics[width=0.36\textwidth]{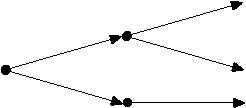}}
  \caption{The network begins as a single vertex and at each iteration one or two new edges are attached to each 'leaf' of the tree, the tree may grow along a single branch as in Fig.\ref{fig:gull} with probability $p$, or split in to two branches as in Fig.\ref{fig:tiger} with probability $1-p$. Figure \ref{fig:mouse} shows what a tree may look like after 2 iterations this particular realization will occur with probability $(1-p)\times(1-p)\times p$. At the far right of the tree all the leaf vertices will, by definition, join to a single busbar to complete the circuit, therefore the network can be seen as a complex combination of resistors in series and in parallel, Fig.\ref{fig:mouse} for example will have total resistance $6/7$ (assuming the resistance across each edge is 1.}
  \label{fig:animals}
\end{figure}

For several years random sequences have been a topic of interest for a number of researchers. While this body of work has been accepted as a branch of statistical physics, the current literature is primarily focused on problems of a purely mathematical conception, namely the idea of a random Fibonacci sequence introduced in \cite{viswanath} and expanded on in \cite{BenNaim}, \cite{Sire} and  \cite{Rodgers}. The natural response to these analyses is to consider areas in applied science where a random sequence may be characteristic of the phenomena being studied, these are disordered systems whose behaviour is non-deterministic in that the state of the system after a short step in time could be any of a number of possibilities (according to certain probabilities), much in an analogous way to a random Fibonacci sequence. One success of statistical mechanics has been the widespread utilization of of complex (random) networks to model naturally occurring phenomena, for this reason random sequences that mimic the properties of random network problems have potential to become a fruitful topic of research. The example considered here extends a number of well studied problems involving networks of electrical resistors, \cite{RandResReview} and \cite{RandomRes} are concerned with the resistance between two sites on a lattice where each edge is a resistor, and \cite{phase} goes further by examining the percolation that occurs when these resistors are overloaded, destroying the corresponding edge and breaking the lattice. On a similar theme, this paper attempts to find the resistance across a particular class of random network where each edge represents a resistor, this paper concerns a theoretical application of the equations associated with electrical resistance, the aim being to find results regarding the total resistance of a random network where each edge represents a resistor, the particular problem chosen has provided an opportunity to show that random sequences can be useful in studying problems in physics.
\newline
The network studied here is grown from a single vertex by the repeated process of appending either one branch or two (with probabilities $p$ and $1-p$ respectively) to those vertices created in the previous iteration (see Fig.\ref{fig:animals}), the tree after $n$ steps is denoted $T_{n}$ [$\equiv T_{n}(p)$]. To simplify the problem all edges are chosen to have a resistance of $1 \Omega$, throughout the paper the units of resistance $\Omega$ will not be displayed. This paper is concerned with the the following question: as a function of $p$, what is the resistance $R_{n}$ [$\equiv R_{n}(p)$] between the root vertex and a busbar connecting all the vertices at the $n^{th}$ level, and what happens when $n\rightarrow\infty$?
The problem is interesting since the equations governing electrical resistance will take a different form depending on whether a given vertex branches in two or not, if it does (Fig.\ref{fig:tiger})then the formula for the resistance across the two parallel edges with resistances $R_{1}$ and $R_{2}$ given by
\begin{equation}
\frac{1}{R_{Total}}=\frac{1}{R_{1}}+\frac{1}{R_{2}}\nonumber
\end{equation}
is used, edges connected in series use the formula $R_{Total}=R_{1}+R_{2}$. The random combination of these equations makes the question of the total resistance difficult to solve, moreover the problem increases rapidly in complexity as the network grows.  \newline
The remainder of this paper describes ways in which these problems are mitigated and approximate solutions are found for $\langle R_{n} \rangle$, its distribution $P_{n}(R)$ as well as the rate of convergence to the mean as $n$ increases. In Section \ref{extreme} the exact solutions for the two special cases, $p=0$ and $p=1$, are presented. In Section \ref{sect} a simplified model is used to approximate $T_{n}(p)$ and the mean and second moment are approximated for general $p$ and large $n$. Section \ref{numer} introduces a method to generate $T_{n}$ accurately and the corresponding numerical results are compared with those of \ref{sect}. A random sequence model is presented in Section \ref{ranseq} from which the convergence towards the mean is obtained as $n$ increases.
\section{$p=0$ and $p=1$}
\label{extreme}
\begin{wrapfigure}{r}{0.5\textwidth}
  \begin{center}
    \label{binary}\includegraphics[width=0.38\textwidth]{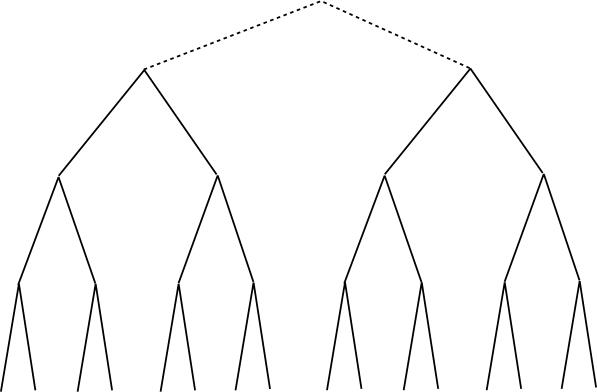}
  \end{center}
  \caption{$T_{n}(0)$ is equivalent to joining together two copies of $T_{n-1}(0)$ by two parallel edges to a new root vertex.}
  \vspace{5mm}
\end{wrapfigure}
At the extreme values, $p=1$ and $p=0$, upper and lower bounds for the $R_{n}(p)$ are easily found: in the first case there is no branching so $T_{n}(1)$ is composed of a line of $n$ edges connected in series; supposing the network grows with $n$, the equation for resistance in series gives
\begin{equation}
R_{n}=R_{n-1}+1 \nonumber
\end{equation}
and so $R_{n}\sim n$ as $n\rightarrow\infty$. In the second case the network branches at every vertex (thus $T_{n}(0)$ is a \emph{complete binary tree}) and so both equations for resistance in series and in parallel are needed.
As illustrated in Fig.\ref{binary}, $T_{n}$ is equivalent to joining two networks, $T_{n-1}$, by two parallel edges from the root vertex to a newly created root vertex (to verify this, observe that $T_{n}$ has $2^{n}$ end points (leaves) and $2\times T_{n-1}$ has $2\times 2^{n-1}$). The consequent resistance equation is
\begin{align}
\frac{1}{R_{n}}=&\frac{1}{R_{n-1}+1}+\frac{1}{R_{n-1}+1} \nonumber \\
\Rightarrow R_{n}=&\frac{R_{n-1}+1}{2}, \nonumber
\end{align}
the solution to this being $R_{n}\rightarrow 1$ as $n\rightarrow \infty$.
\newpage
\section{A simplified model to approximate $R_{n}$}
\label{sect}
\begin{wrapfigure}{r}{0.5\textwidth}
  \begin{center}
    \label{dupl}\includegraphics[width=0.38\textwidth]{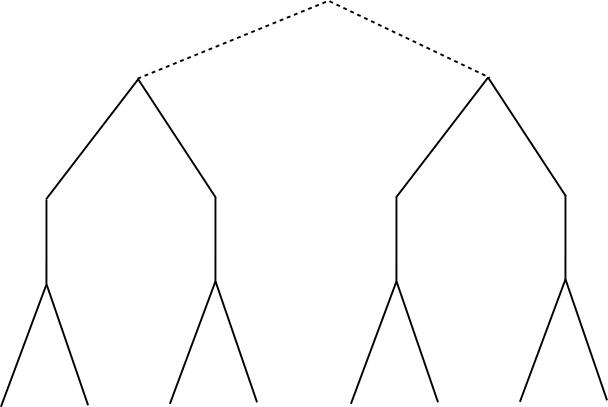}
  \end{center}
  \caption{In the simplified model at each iteration the network is either extended by an edge from the root vertex or joined with a duplicate of itself as shown.}
\end{wrapfigure}
When $0<p<1$ one wishes to find the mean $\langle R_{n}\rangle$ as a function of $p$ as well as higher moments and also the distribution $P_{n}(R)$, since this is not easily obtained a simplified network $T_{n}^{*}$ is studied. In this section, for the resistance $R$ of $T_{n}^{*}$ when $n\rightarrow\infty$ the first and second moments are found and the distribution $P(R)$ is expressed in two different forms.

\begin{figure}
  \begin{center}
    \includegraphics[width=0.6\textwidth]{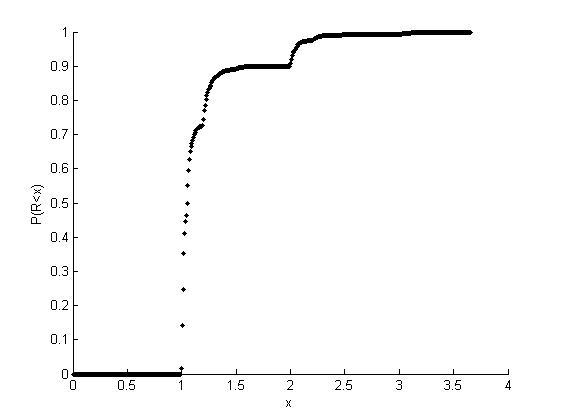}
  \end{center}
  \caption{Cumulative probability distribution plotted with the output of $10^{4}$ realizations of the simplified model (Model I), $n=10^{6}$ and $p=0.1$.}
  \label{smallp}
\end{figure}

With probability $p$, $T_{n}^{*}$ is constructed by joining the root vertex of $T_{n-1}^{*}$ to a new vertex or, with probability $1-p$, $T_{n}^{*}$ is found by joining two duplicates of $T_{n-1}^{*}$ to a new vertex (see Fig.\ref{dupl}). Since this network retains the same proportion of split branches as $T_{n}$ one expects it to be a close approximation. The corresponding resistances are given by
\begin{equation}
R_{n} = \left\{
  \begin{array}{l l}
  \label{problem}
    R_{n-1}+1 & \quad \text{with probability $p$} \\
    \frac{R_{n-1}+1}{2} & \quad \text{with probability $1-p$}\\
  \end{array} \right.
\end{equation}
One can immediately write down a recursive formula for the average,
\begin{align}
\label{steady}
\langle R_{n}\rangle=\frac{1+p}{2}\left[\langle R_{n-1}\rangle+1\right],
\end{align}
which indicates the steady state value for which $R_{n}$ converges to: $\langle R\rangle=(1+p)/(1-p)$.
The distribution of $R$ at iteration $n$ obeys
\begin{equation}
P_{n}(R)= \int P_{n-1}(R')dR'\left[(1-p)\delta\left(\frac{R'+1}{2}-R\right)+p\delta\left(R'+1-R\right)\right] \nonumber
\end{equation}
where $\delta(x)$ is the Dirac delta function. This to simplifies to
\begin{align}
 \label{firstProb}
P_{n}(R)=2(1-p)P_{n-1}(2R-1)+pP_{n-1}(R-1).
\end{align}
For the second moment, following directly from Eq.(\ref{firstProb}),
\begin{align}
\langle R_{n}^{2}\rangle=\int_{1}^{\infty}R^{2}P_{n}(R)dR= 2(1-p)\int_{1}^{\infty}R^{2}P_{n-1}(2R-1)dR+p\int_{1}^{\infty}R^{2}P_{n-1}(R-1)dR \nonumber
\end{align}
is solved with changes of variable, $u=R-1$ and $v=2R-1$, and the knowledge that for any natural number $n$, $\int_{0}^{1}P_{n}(R)dR=0$ and $\int_{1}^{\infty}P_{n}(R)dR=1$ which follows from Eq.(\ref{problem}). The resulting recursive formula is
\begin{equation}
\langle R_{n}^{2}\rangle=\frac{1+3p}{4}(\langle R_{n-1}^{2}\rangle+2\langle R_{n-1}\rangle+1),
\end{equation}
as $n\rightarrow \infty$ the second moment is found to be $\langle R^{2}\rangle=(3+10p+3p^{2})/3(1-p)^2$.
\newline
Additionally, $P_{n}(R)$ converges to an invariant distribution as $n$ increases to infinity, from Eq.(\ref{firstProb}) the invariant distribution satisfies
\begin{equation}
\label{invariant}
P(R)=2(1-p)P(2R-1)+pP(R-1).
\end{equation}
Using $\tilde{P}(k)$ to denote the Laplace transform of $P(R)$, the solution of Eq.(\ref{invariant}) when transformed,
\begin{equation}
\tilde{P}(k)=\mathcal{L} (P(R))=2(1-p)\int_{1}^{\infty}P(2R-1)e^{-kR}dR+p\int_{1}^{\infty}P(R-1)e^{-kR}dR, \nonumber
\end{equation}
simplifies to the recursive equation
\begin{align}
\tilde{P}(k)=&\frac{(1-p)e^{k/2}}{1-pe^{-k}}\tilde{P}(k/2) \nonumber\\
=&\frac{(1-p)e^{-k/2}}{1-pe^{-k}}\frac{(1-p)e^{-k/4}}{1-pe^{-k/2}}\tilde{P}(k/4) \nonumber\\
\vdots& \nonumber\\
\label{denom}
=&\prod_{r=0}^{\infty}\frac{(1-p)e^{-k/(2^{r+1})}}{1-pe^{-k/(2^{r})}}
\tilde{P}(k)=e^{-k}\prod_{r=0}^{\infty}\frac{1-p}{1-pe^{-k/2^{r}}}
\end{align}
The inverse Laplace transform will recover an expression for $P(R)$, this is described as
\begin{align}
P(R)=&\frac{1}{2\pi\emph{i}}\lim_{T\rightarrow\infty}\int_{\gamma-\emph{i}T}^{\gamma+\emph{i}T}\tilde{P}(k)e^{Rk}dk \\
=&\text{ the sum of the residues of }\tilde{P}(k)e^{Rk}.
\end{align}
These residues lie at the points on the complex plane where the denominator in Eq.(\ref{denom}) is equal to zero, i.e for each root $k_{s}$ ($s=0,1,2,...$), $k_{s}=2^{s}\log(p)$.
Calculating and summing these residues yields
\begin{equation}
P(R)=\sum_{s=0}^{\infty}2^{s}p^{2^{s}R}\frac{1-p}{\sqrt{p}}\prod_{r\neq s}^{\infty}\frac{(1-p)p^{-2^{s-r-1}}}{1-p^{1-2^{s-r}}}. \nonumber
\end{equation}
\newline
Using the expansion $1/(1-X)=1+X+X^{2}+...$ with $X=pe^{-k/2^{r}}$, Eq.(\ref{denom}) can be written
\begin{align}
\tilde{P}(k)=e^{-k}\prod_{r=0}^{\infty}(1-p+pe^{-k/2^{r}}-p^{2}e^{-k/2^{r}}+p^{2}e^{-k/2^{r-1}}-...) \nonumber
\end{align}
Focusing only on terms up to and including multiples of $p^{2}$, multiplying out the brackets and recalling the translation property of $\delta(x)$,
\begin{align}
&\tilde{P}(k)=e^{-k}+p\sum_{r=0}^{\infty}(e^{-k(1+1/2^{r})}-e^{-k}) \nonumber\\
&+p^{2}\left(\sum_{r=0}^{\infty}(e^{-k(1+1/2^{r-1})}-e^{-k(1+1/2^{r})})+\sum_{i=0}^{\infty}\sum_{j=0}^{\infty}(e^{-k(1+1/2^{i}+1/2^{j})}-e^{-k(1+1/2^{i})}-e^{-k(1+1/2^{j})}+e^{-k})\right) \nonumber
\end{align}
can be expressed as
\begin{align}
\tilde{P}(k)&=\int e^{-kR}\delta(R-1)dR+p\sum_{r=0}^{\infty}\int e^{-kR}\delta(R-1-1/2^{r})dR-...\nonumber\\
&=\int e^{-kR}P(R)dR\\ \nonumber
\text{where }
\label{secOrd}
P(R)&\approx\delta(R-1)+p\sum_{r=0}^{\infty}[\delta(R-1-1/2^{r})-\delta(R-1)]+p^{2}\sum_{r=0}^{\infty}[\delta(R-1-1/2^{r-1})-\delta(R-1-1/2^{r})]\\
&+p^{2}\sum_{i=0}^{\infty}\sum_{j=0}^{\infty}[\delta(R-1-1/2^{i}-1/2^{j})-\delta(R-1-1/2^{i})-\delta(R-1-1/2^{j})+\delta(R-1)].
\end{align}
As values of $p$  go towards zero, the values identified by the delta functions in the above expression constitute an increasingly significant proportion of $P(R)$. This can be seen in Fig.\ref{smallp} with the largest probability occurring at $R=1$, corresponding to the first order term as well as lower order terms, other notable values of $R$ correspond to the values identified by the delta functions that are multiplied by $p^{2}$ in Eq.(\ref{secOrd}), $2$, $1.5$, $1.25$, etc..
\section{Comparison of the resistances of $T^{*}_{n}$ and $T_{n}$: numerical results}
\label{numer}
\begin{figure}
  \begin{center}
    \includegraphics[width=0.6\textwidth]{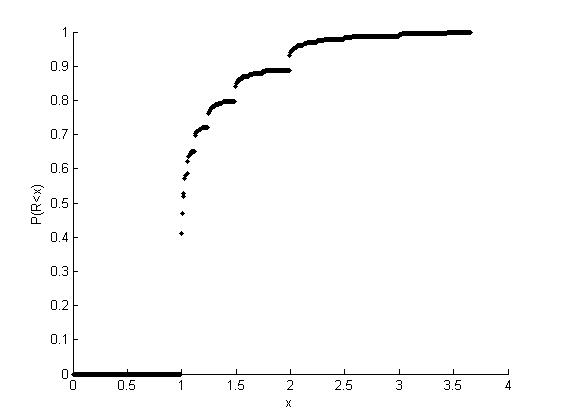}
  \end{center}
  \caption{Cumulative probability distribution plotted with the output of $10^{4}$ realizations of the accurate model, $n=10^{6}$ and $p=0.1$.}
  \label{cumul2}
\end{figure}

In this section the results of Section \ref{sect} are compared with numerical results generated by simulation of the dynamical system Eq.(\ref{problem}) (Table \ref{table}), additionally a second system is introduced which accurately reproduces the behaviour of $R_{n}$ for the original network $T_{n}$. The similarity of the simplified model with the accurate representation is then shown by a comparison of each models mean (Fig. \ref{compare} and Table \ref{table}), variance (Table \ref{table}) and distribution (Figs. \ref{smallp} and \ref{cumul2}) generated numerically.
\subsection{Constructing $T_{n}$}
\begin{figure}
  \begin{center}
    \includegraphics[width=0.6\textwidth]{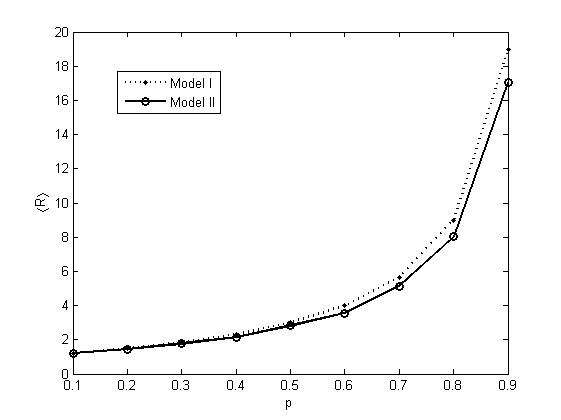}
  \end{center}
  \caption{Comparison of the accurate model and the simplified model ($n=10^{6}$).}
  \label{compare}
\end{figure}

With probability $p$, $T_{n}$ is constructed by joining the root vertex of $T_{n-1}$ to a new vertex or, with probability $1-p$, $T_{n}$ is found by joining two trees $T_{a}$ and $T_{b}$ to a single root vertex, where both $T_{a}$ and $T_{b}$ are possible realizations of $T_{n-1}$. The resistance of $T_{n}$ is then given by
\begin{equation}
R_{n} = \left\{
  \begin{array}{l l}
    R_{n-1}+1 & \quad \text{with probability $p$} \\
    \cfrac{1}{1/(R_{a}+1)+1/(R_{b}+1)} & \quad \text{with probability $1-p$}\\
  \end{array} \right. \vspace{3mm}
\end{equation}
where $R_{a}$ and $R_{b}$ are distributed according to $P_{n-1}(R)$. Comparisons are shown in table \ref{table} and Fig.\ref{compare}, for the simple model it was shown that the mean converges to $(1+p)/(1-p)$ and the sum of the squares converges to $(3+10p+3p^2)/3(1-p)^{2}$ from which the variance is calculated.\newline
\begin{table}
\caption{A comparison is made between the simplified model and the accurate model, here the results for the mean and variance predicted in \ref{sect} are shown next to numerical results obtained from $10^{6}$ realizations. Numerical results of the accurate model are also shown here and in Fig. \ref{compare}, illustrating the accuracy of the simplified model as and approximation to the accurate one.}
\begin{center}
\begin{tabular}{ c  c  c  c  c  c  c}
\hline \hline
p & \multicolumn{2}{c}{Predicted} & \multicolumn{2}{c}{Model I (Simplified)} & \multicolumn{2}{c}{Model II (Accurate)} \\
&Average & Variance & Average & Variance & Average & Variance \\
\hline
0.1&1.22222&0.164618&1.22266&0.173222&1.19601&0.132801\\
0.2&1.5&0.416667&1.49595&0.422622&1.43467&0.337122\\
0.3&1.85714&0.816326&1.86260&0.850533&1.74874&0.666383\\
0.4&2.33333&1.48148&2.32556&1.45536&2.17774&1.19880\\
0.5&3&2.66667&3.00448&2.68774&2.72414&2.10841\\
0.6&4&5&3.96578&4.77015&3.60162&3.92301\\
0.7&5.66667&10.3704&5.67442&10.4559&5.04931&8.19541\\
0.8&9&26.6667&8.98651&26.7466&8.36457&23.7019\\
0.9&19&120&18.8040&116.624&17.3566&104.912\\
\hline \hline
\end{tabular}
\label{table}
\end{center}
\end{table}

\section{Using random sequences to predict the convergence of $R_{n}$}
\label{ranseq}
To obtain the rate at which $\langle R_{n}\rangle$ converges to the mean a third model is considered in this section. Expressing $\langle R_{n}\rangle$ as a recurrence relation it is shown to be equivalent to a well known family of orthogonal polynomials, these polynomials are expressed as an integral which is solved to retrieve the average as a function of $p$ and $n$ when $n$ is large.\newline
In this model the tree $T_{n}$ is either extended by an edge from the root vertex as before (with probability $p$) or two duplicates of $T_{q}$ are connected to a newly created root, where $q$ is a randomly selected integer from $[0,n-1]$. The resistance is then given by following the random sequence
\begin{equation}
R_{n} = \left\{
  \begin{array}{l l}
    R_{n-1}+1 & \quad \text{with probability $p$} \\
    \cfrac{R_{q}+1}{2} & \quad \text{with probability $1-p$ and $q \in [0,n-1]$}\\
  \end{array} \right.
\end{equation}
If the $T_{q}$ are chosen with equal probability then for large $n$ one would expect the distribution of $R_{q}$ to approach that of $R_{n}$, this specifically describes a system of either attaching a single edge to the root vertex of $T_{n}$ (with probability $p$) or selecting a previous $T_{q}$ and attaching it to its own duplicate (with probability $1-p$). Letting $Q_{n}(q)$ be the probability that the value $q\in[0,n-1]$ is chosen, the distribution of $R_{n}$ obeys the integral equation
\begin{align}
P_{n}(R)=& \int(1-p)\sum_{q=0}^{n-1}Q_{n}(q)P_{q}(R')\delta\left(R-\frac{R'+1}{2}\right)dR' \nonumber \\
&+\int pP_{n-1}(R)\delta(R-(R'+1))dR' \nonumber
\end{align}
which reduces to
\begin{equation}
P_{n}(R)=2(1-p)\sum_{q=0}^{n-1}Q_{n}(q)P_{q}(2R-1)+pP_{n-1}(R-1) \nonumber
\end{equation}
From this the average $A_{n}=\langle R_{n}\rangle=\int RP_{n}(R)dR$ is found to obey
\begin{equation}
\label{AvRec}
A_{n}=(1-p)\sum_{q=0}^{n-1}Q_{n}(q)\left(\frac{A_{q}+1}{2}\right)+p(A_{n-1}+1).
\end{equation}
A similar argument to the following can be found in \cite{Rodgers}, the distribution $Q_{n}(q)$ can be written as $Q_{n}(q)=Q(q)/b_{n}$ where
\begin{equation}
\label{bsum}
b_{n}=\sum_{q=0}^{n-1}Q(q).
\end{equation}
Then Eq.(\ref{AvRec}) becomes
\begin{equation}
\label{avSum}
A_{n}=\frac{(1-p)}{b_{n}}\sum_{q=0}^{n-1}Q(q)\left(\frac{A_{q}+1}{2}\right)+p(A_{n-1}+1)
\end{equation}
Subtracting Eq.(\ref{avSum}) from the equivalent equation for $A_{n+1}$, and observing from Eq.(\ref{bsum}) that $Q(n)=b_{n+1}-b_{n}$, it is found that
\begin{equation}\label{simp},
2b_{n+1}A_{n+1}+2pb_{n}A_{n-1}=(p+1)[(b_{n+1}+b_{n})A_{n}+(b_{n+1}-b_{n})].
\end{equation}
In the case where the previous $T_{n}$ are chosen with equal probability, $Q_{n}(q)=1/n$ for all $q\in [0,n-1]$, $b_{n}=n$ ($b_{n+1}=n+1$, obviously) and Eq.(\ref{simp}) becomes
\begin{equation}
2(n+1)A_{n+1}+2pnA_{n-1}=(p+1)[(2n+1)A_{n}+1]
\end{equation}
equivalently
\begin{equation}
\label{withA}
nA_{n}=\frac{p+1}{2}[(2n-1)A_{n-1}+1]-p(n-1)A_{n-2}.
\end{equation}
A solution can be obtained with the help of some known results in orthogonal polynomials, this is possible by first observing that the transformation
\begin{equation}\label{trans}
A_{n}=\frac{1+p}{1-p}+p^{\frac{n}{2}}B_{n}
\end{equation}
when substituted into Eq.(\ref{withA}) leaves
\begin{equation}
nB_{n}=\frac{p+1}{2\sqrt{p}}(2n-1)B_{n-1}-(n-1)B_{n-2},
\end{equation}
the recursion relation for the Legendre polynomials $B_{n}=P_{n-1}(x)$ at $x=(p+1)/2\sqrt{p}$ \cite{absteg}.
Given in \cite{absteg}, the Gegenbauer polynomials, which obey the integral form
\begin{equation}\label{Geg}
C^{(\alpha)}_{n}(x)=\frac{2^{(1-2\alpha)}\Gamma(n+2\alpha)}{n![\Gamma(\alpha)]^{2}}\int_{0}^{\pi}[x+\sqrt{x^{2}-1}\text{ cos }\phi]^{n}(\text{ sin }\phi)^{2\alpha-1}d\phi,
\end{equation}
become the Legendre polynomials at the value $\alpha=1/2$, so Eq.(\ref{Geg}) reduces to the much simpler
\begin{equation}\label{leg}
P_{n}(x)=\frac{1}{\pi}\int_{0}^{\pi}[x+\sqrt{x^{2}-1}\text{ cos }\phi]^{n}d\phi.
\end{equation}
This can be easily solved using Laplace's method as it can be expressed in the form
\begin{equation}
P_{n}(x)=\frac{1}{\pi}\int_{0}^{\pi}\text{ exp}\{nf(\phi)\}d\phi \nonumber
\end{equation}
with
\begin{align}
f(\phi)=&\text{ log}[x+\sqrt{x^{2}-1}\text{ cos }\phi]\nonumber\\
f'(\phi)=&-\frac{\sqrt{x^{2}-1}\text{ sin }\phi}{x+\sqrt{x^{2}-1}\text{ cos }\phi}\label{firstd}\\
f''(\phi)=&-\frac{\sqrt{x^{2}-1}\text{ cos }\phi}{x+\sqrt{x^{2}-1}\text{ cos }\phi}+\frac{(x^{2}-1)\text{ sin}^{2}\phi}{[x+\sqrt{x^{2}-1}\text{ cos }\phi]^{2}}.\label{second}
\end{align}
From Eq.(\ref{firstd}) it can be seen that stationary points of $f$ exist at $\phi=0$ and $\phi=\pi$, putting these into Eq.(\ref{second}) yields
\begin{align}
\label{fzero}
f''(0)=&-\frac{\sqrt{x^{2}-1}}{x+\sqrt{x^{2}-1}}\\
\text{and }f''(\pi)=&\frac{\sqrt{x^{2}-1}}{x+\sqrt{x^{2}-1}}
\end{align}
Using Laplace's method on the integral in Eq.(\ref{leg}), as $n\rightarrow\infty$
\begin{align}
P_{n}(x)&\sim\frac{1}{\pi}\frac{1}{2}\sqrt{\frac{2\pi}{n|f''(\phi_{0})|}}\text{ exp}\{nf(\phi_{0})\} \nonumber\\
&=\frac{1}{2\pi}\sqrt{\frac{2\pi}{n}}\sqrt{\frac{x+\sqrt{x^{2}-1}}{\sqrt{x^{2}-1}}}(x+\sqrt{x^{2}-1})^{n} \nonumber\\
\Rightarrow P_{n}\left(\frac{1+p}{2\sqrt{p}}\right)&\approx\frac{1}{\sqrt{n\pi}}\sqrt{\frac{1}{1-p}}p^{-n/2}. \nonumber
\end{align}
Note that only half of the value is taken since the maximum is on the boundary of the integral. Relating this back to the formula for the average value of the random sequence [Eq.(\ref{trans})],
\begin{equation}
A_{n+1}=\frac{1+p}{1-p}+p^{\frac{n+1}{2}}P_{n}\left(\frac{p+1}{2\sqrt{p}}\right),\nonumber\\
\end{equation}
the resulting equation showing the rate of convergence
\begin{equation}
\label{final}
A_{n+1}=\frac{1+p}{1-p}+\frac{1}{\sqrt{\pi n}}\sqrt{\frac{p}{1-p}},
\end{equation}
also retrieves the value for $\langle R_{n}\rangle$ for large $n$ seen in Section \ref{sect}.

\begin{figure}
  \begin{center}
    \includegraphics[width=0.8\textwidth]{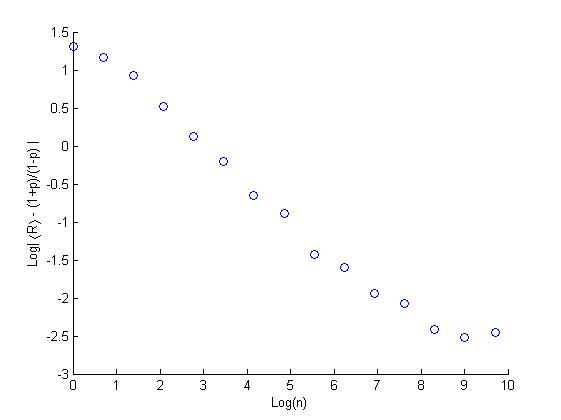}
  \end{center}
  \caption{Convergence of random sequence representation of $\langle R_{n} \rangle$, here $p=0.7$. The gradient in this plot agrees with  the predicted result in Eq.(\ref{final}) that the convergence to the mean behaves as $n^{-1/2}$.}
\end{figure}
\section{Summary}
A class of random resistor networks has been introduced with a tree-like structure characterized by a single parameter $p$. By constructing similar yet simplified tree networks that retained the important property of the proportion of branching points to non-branching points, approximations were made to the resistance of the network with a slight loss of accuracy. For a growing network it was established for the approximation that when $p=1$ the resistance diverges but for all other values of $p$ the resistance converges to $(1+p)/(1-p)$ with $|\langle R_{n}\rangle-(1+p)/(1-p)|\sim n^{-1/2}$. It was revealed that the structure of the probability distribution of $R_{n}$ when $n$ is large is intricate although as $p$ decreases certain values begin to dominate the distribution.
\section*{Acknowledgments}
This research was supported by EPSRC.

\bibliography{bibfile}

\begin{thebibliography}{10}

\bibitem{viswanath}
D.~Viswanath.
\newblock Mathematics of Computation \textbf{volume~69}, 231 (2000) pp. 1131.

\bibitem{BenNaim}
E.~Ben-Naim and P.~L. Krapivsky.
\newblock Journal of Physics A: Mathematical and General \textbf{volume~35}, 41
  (2002) L557.

\bibitem{Sire}
C.~Sire and P.~L. Krapivsky.
\newblock Journal of Physics A: Mathematical and General \textbf{volume~34}, 42
  (2001) 9065.

\bibitem{Rodgers}
I.~Krasikov, G.~J. Rodgers, and C.~E. Tripp.
\newblock Journal of Physics A: Mathematical and General \textbf{volume~37}, 6
  (2004) 2365.

\bibitem{RandResReview}
J.~H. Asad, A.~Sakaji, R.~S. Hijjawi, and J.~M. Khalifeh.
\newblock The European Physical Journal B - Condensed Matter and Complex
  Systems \textbf{volume~52} (2006) 365.
\newblock 10.1140/epjb/e2006-00311-x.

\bibitem{RandomRes}
J.~H. Asad, R.~S. Hijjawi, A.~Sakaj, and J.~M. Khalifeh.
\newblock International Journal of Theoretical Physics \textbf{volume~44}
  (2005) 471.
\newblock 10.1007/s10773-005-3977-6.

\bibitem{phase}
D.~Stauffer and A.~A.
\newblock \emph{Introduction to Percolation Theory}.
\newblock 2 edition (Taylor and Francis, London, Washington, DC, 1991).

\bibitem{absteg}
M.~Abramowitz and I.~Stegun.
\newblock \emph{Handbook of Mathematical Functions} (Dover, New York, 1975).

\bibitem{Makover}
E.~Makover and J.~McGowan.
\newblock Journal of Number Theory \textbf{volume 121}, 1 (2006) 40 .

\bibitem{Dux}
P.~M. Duxbury, P.~L. Leath, and P.~D. Beale.
\newblock Phys. Rev. B \textbf{volume~36}, 1 (1987) 367.

\end{thebibliography}
\bibliographystyle{tues2}

\pagestyle{plain}
\end{document}